\documentclass[article,aps,prd,12pt,notitlepage]{revtex4-1}

\usepackage[utf8]{inputenc}
\usepackage[T1]{fontenc}

\usepackage{bm}
\usepackage{amssymb}
\usepackage{graphicx,xcolor}
\usepackage{csquotes}
\usepackage{amsmath,graphicx}
\begin{document}
\def\la{{\langle}}
\def\u{\hat U}
\def\A{\mathcal A}
\def\PP{\mathcal P}
\def\B{\hat B}
\def\C{\hat C}
\def\D{\hat D}
\def\Om{\Omega}
\def\up{\uparrow}
\def\do{\downarrow}
\def\ep{\epsilon}
\def\fb{\overline F}
\def\pia{\hat \pi_A}
\def\wb{\overline W}
\def\nl{\newline}
\def\h{\hat H}
\def\G{\Gamma}
\def\lm{\lambda}
\def\lmu{\underline\lambda}
\def\q{\quad}
\def\t{\tau}
\def\om{\omega}
\def\s{\mathcal{S}}
\def\r{\color{red}}
\def\g{\color{green}}
\def\b{\color{blue}}
\def\n{\\ \nonumber}
\def\ra{{\rangle}}
\def\Ep{{\mathcal{E}}}
\def\E{E_r}
\def\Ee{{\epsilon}}
\def\a{{\hat a}}
\def\sx{{\hat \sigma_x}}
\def\sy{{\hat \sigma_y}}
\def\sz{{\hat \sigma_z}}
\def\h{\hat{H}}
\def\ha{\hat{H}_A}
\def\ua{\hat U_A}
\def\uu{\hat u}
\def\rh{{\rho}}
\def\hh{{\mathcal h}}
\def\e{\enquote}
\def\d{{e}}
\def\nn{n_1}
\def\nnn{\overline{n}_1}
\def\nm{\overline{n}}
\title{Timing the moment when atom decays \begin {center} \unskip (and Schr\"odinger's  cat dies) \end{center}}

 \date\today
 
%
%
\author {D. Sokolovski$^{a,c,d,*}$}
\author {A. Uranga$^{b}$}
\author {E. Akhmatskaya$^{b,c}$}
\affiliation{$^a$ Departmento de Qu\'imica-F\'isica, Universidad del Pa\' is Vasco, UPV/EHU, 48940, Leioa, Spain}
\affiliation{$^b$ Basque Center for Applied Mathematics (BCAM),\\ Alameda de Mazarredo 14, 48009, Bilbao, Spain}
\affiliation{$^c$ IKERBASQUE, Basque Foundation for Science, Plaza Euskadi 5, 48009, Bilbao, Spain}
\affiliation{$^d$ EHU Quantum Center, Universidad del Pa\' is Vasco, UPV/EHU, 48940, Leioa, Spain}
\email{Corresponding author. Email: dgsokol15@gmail.com}
\begin{abstract}
	\textbf{\abstractname.}
	We 
	\textcolor{black}{propose detecting}
	the  moment an atom emits a photon
	by means of a nearly classical macroscopic clock and discuss its viability. 
	It is shown that what happens in such a measurement depends 
	on the relation between 
	\textcolor{black}{the clock's} accuracy 
	and the width 
	of the energy range available to the photon.
	Implications of the analysis for the long standing Schr\"odinger's cat problem are
	\textcolor{black}{reported.} 
\end{abstract}

%
%
\maketitle
%
%
\vspace{0.1cm}
\section*{Introduction }
In non-relativistic quantum mechanics, time is a mere parameter, quite distinct from the dynamical variables such as positions and momenta, 
conveniently represented by Hermitian operators. This often complicates the queries, easily answered in the classical context 
(a good overview has been given in refs. \cite{TQ1} and \cite{TQ2}). When does a quantum particle arrive \textcolor{black} {at} a given location (see Egusquiza, Muga and Baute in \cite{TQ1} and Galapon in \cite{TQ2}? How much time does a tunnelling particle spend in the barrier (see, e.g., \cite{Swiss}, \cite{DSnat})?  
How long does a quantum jump take (see Schulman, in \cite{Schul} and Refs. therein)? These questions continue to cause controversy, 
and here we add one more to the list.
\newline
If an atom, initially in an excited state, emits a photon and is  later found in its ground state, 
when exactly did the transition take place? If the decay sets off a chain of events leading to the death of a cat \cite{Cat}, 
how long ago did the cat die? This is another general problem in 
elementary quantum mechanics, and below we will address it, using the simplest model available.
\newline 
\subsection*{A meaningful question?}
Does it make sense to talk about the moment the atom decayed? Not always.
Decay of a metastable state is often described by a model  \cite{Gurv}
where a discrete state $|e\ra$, corresponding to an excited \e {atom} with energy $E_\d$, is connected to \e{reservoir} 
states $\{|\E\ra\}$, representing  an \e{atom} in its ground state, $E_g=0$, plus an emitted \e{photon} with energy  $\E$.
The corresponding Hamiltonian takes the form,
\begin{eqnarray}\label{1b}
\h =\h_0 +\hat V,\q
\h_0=|\d\ra E_\d\la \d| + \sum_r |\E\ra \E\la \E|,\n
\hat V=
\sum_r  \Omega(\E)\left (|\E\ra \la \d| + |\d\ra \la \E|\right), \q\q\q
\end{eqnarray}
where $\Omega(\E)$ is the matrix element responsible for the transitions between the system's discrete and continuum states, 
i.e., for the decay of the excited atom. 
In the continuum limit, whenever the final probabilities are added up, one can replace the sum $\sum_r$ 
by an integral $\int \rho(\E) d\E$, where $ \rho(\E)$ is the density of the reservoir states \cite{Gurv}. 
\newline
After preparing an atom in its excited state, and waiting for $t$ seconds, 
one can find a photon with an energy $\E$.
Expanding the transition amplitude $\la \E|\exp(-i\h t)|\d\ra$ in powers of $\hat V$ reveals a variety
of scenarios where the photon, emitted for the first time at $\tau_{\mathrm{first}}$
is re-absorbed and re-emitted until settling down into its final state $|E_r\ra$ at some $\tau_{\mathrm{last}}$.
Thus, the emission process may not occur via a single transition to the ground state, but can have a finite duration $\tau_{\mathrm{last}}-\tau_{\mathrm{first}}$. Measuring even \e{the first passage time} $\tau_{\mathrm{first}}$ presents considerable 
difficulties  \cite{DSpit}, and we do not know if $\tau_{\mathrm{last}}-\tau_{\mathrm{first}}$ can be measured at all.
\newline
A helpful exception is the first order transition in the weak coupling limit, which does indeed occur via a single jump, 
\begin{eqnarray}\label{2b}
\la \E|\exp(-i\h t)|\d\ra =-i\Om(\E) \int_0^t d\t \exp[-i\E(t-\t)]\exp(-iE_\d \t)+ \mathcal{O}(V^3),
\end{eqnarray}
yet  the jump's  precise moment remains indeterminate due to the Uncertainty Principle \cite{FeynL}.
One way to pinpoint the time of transition is to subject the atom to frequent observations
every $\delta t =t/K$, $K>>1$.
This, however, is known to lead to the Zeno effect which quenches the transition, whose rate changes
from the one given by the Fermi's golden rule \cite{Fermi} for \textcolor{black}{an }unobserved atom, $\G_{Fermi} =2\pi|\Om(E_\d)|^2\rh(E_\d)$ to
$\G_{\delta t} \approx \delta t\times  \int_{-\infty}^\infty dE \rh(E)|\Om(E)|^2$, which vanishes as $\delta t \to 0$
(see, e.g., \cite{Schul}).
\newline 
\textcolor{black}{Yet, there} is 
a case where the transition proceeds via a single jump, and the Zeno effect does not occur. 
We will discuss it next. 
\section*{Results and Discussion}
\subsection*{The wide band (Markovian) case}

In the Markovian  (wide band)  approximation \cite{Gurv},
both $\Om(E_r)$ and  $\rh(\E)$, are taken to constant, 
very small and very big respectively, i.e.
$\Om\to 0$,  $\rh\to \infty$, 
in such a manner that
a product $\rh\Om^2$
remains finite, 
\begin{eqnarray}\label{1c}
2\pi\rh \Om^2\equiv \G < \infty.
\end{eqnarray}
The model admits an exact solution for any $\Gamma$, and there is no need 
to limit oneself to the first order approximation (\ref{3c}).
\textcolor{black}{The amplitudes of the four possible processes} are given by  \cite{Gurv}: 
\begin{eqnarray}\label{2c}
\la \d |\exp(-i\h t)|\d\ra= \exp(-iE_\d t -\G t/2),\q\q\q\q\q\q\q\q\q\q\q\q\q\q\q\q\q
\end{eqnarray}
\begin{eqnarray}\label{3c}
\la \E |\exp(-i\h t)|\d\ra 
=-i\Om \int_0^t dt' \exp[-i\E(t-t')]\exp(-iE_\d t'-\G t'/2),\q\q\q\q
\end{eqnarray}
\begin{eqnarray}\label{2ca}
\la \d |\exp(-i\h t)|\E\ra=0 ,\q \text{since}\q \Om \to 0, \q\q\q\q\q\q\q\q\q\q\q\q\q\q\q\q\q\q
\end{eqnarray}
\begin{eqnarray}\label{2cb}
\la E_{r'} |\exp(-i\h t)|\E\ra=\exp(-i\E t)\delta_{rr'} ,\q \text{since}\q \la E_{r'}| \h|\E\ra=\E\delta_{rr'}. \q\q\q\q\q
\end{eqnarray}
By (\ref{2c}),  atom's  decay is exponential at all times, and  by (\ref{3c}), the energy distribution of the emitted photons is Lorentzian  
\begin{eqnarray}\label{5c}
P(E_r\gets \d, t\to \infty)=\frac{\rho\Om^2}{(\E-E_\d)^2+\G^2/4}.  \q\q\q\q\q\q\q\q
\end{eqnarray}
Further helpful to our purpose is the fact that, according to Eqs.(\ref{3c}) and (\ref{2ca}), the atom can emit a photon only once, 
and never re-absorbs it afterwords. The moment of transition can,
therefore, be defined at least in terms of the virtual scenarios available to the system. 
With the purely exponential decay in Eq.(\ref{2c}) frequent checks of the atom's state
do not  affect the decay rate, which stays the same with or without such checks [hence, the adjective {\it Markovian},
$P^{\textrm{decay}}_M( t) = 1- \exp(-\G t)=1- [\exp(-\G t/K)]^K= P^{\textrm{decay}}_{\delta t}(t)$]. 
Even so, destruction of coherence between the moments of emission in Eq.(\ref{3c}) must change something akin to 
the interference pattern in a double slit experiment. Below we will show that it is the energy spectrum of the emitted photons
(\ref{5c}) that is affected by the measurement's accuracy.
\subsection*{A quantum hourglass and its macroscopic limit}
\begin{center}
\begin{figure}[ht]
	\includegraphics[angle=0,width=14cm]{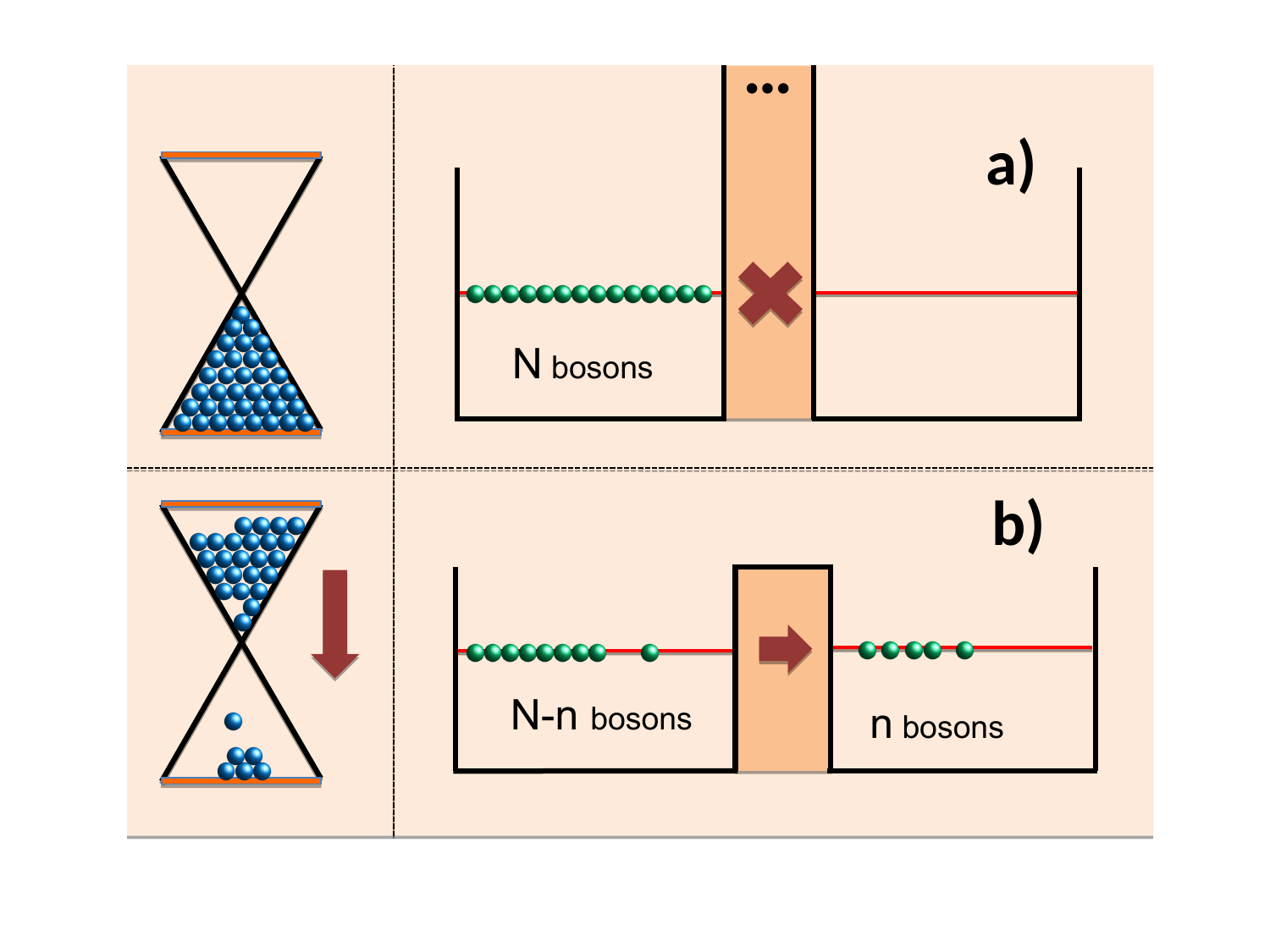}
	\caption{A classical hourglass (left), and its quantum version (right).
		a) With the barrier closed (the clock is switched off) the bosons remain in the left well.
		b) If the barrier is down (the clock is switched on), the number of bosons escaping into the right well allows one to estimate the elapsed time.}
	\label{plot:Fig1}
\end{figure}
\end{center}

Suppose Alice the experimenter, does not wish to subject the system to frequent checks, and prefers 
instead to have, at the end of the experiment, a single record of the moment the atom decayed.
For this purpose, she might consider a clock which stops at the moment the atom leaves its excited state. 
The clock could be an hourglass, in which case the number of the sand grains escaped, 
would tell Alice the time of the event. A quantum analogue of an hourglass is not difficult to find. 
Alice could use  an array of identically polarised distinguishable spins precessing in a magnetic field,
and estimate the elapsed time by counting the spins which have been flipped.
Alternatively, Alice can employ a large number of non-interacting bosonic atoms, $N>>1$, initially in the left well of
a symmetric double well potential (see Fig.\ref{plot:Fig1}). 
The clock's Hamiltonian given by 
[$\a^+_{R(L)}$ creates a boson in the right  (R),  or left (L) well, and $\om$ is the hopping matrix element]
\begin{eqnarray}\label{n1}
\h^{\textrm{clock}}=\om [\a^+_R \a_L+\a^+_L \a_R ], 
\end{eqnarray}
and the amplitude of finding $n$ bosons in the right well is easily found to be 
\begin{eqnarray}\label{n2}
A^{\textrm{clock}}_{\textrm{Bose}} (n\gets 0,t)= (-i)^n\sqrt {C^N_n p^n\textcolor{black}{(t)}(1-p\textcolor{black}{(t)})^{N-n}}, \q p(t) \equiv \sin^2(\om t),
\end{eqnarray}
where $C^N_n=\frac{N!}{n!(N-n)!}$ is the binomial coefficient.
\newline 
Alice  can choose $\om t <<1$, so that the Rabi period of a single boson is very large,
and have a practically irreversible flow of bosons from left to right.  She can also assure\textcolor{black}{, by making $N$ very large,} that the mean number of atoms in the right well is also large (except perhaps at very short times), $\overline n(t)\equiv p(t) N >>1$. Under these conditions, binomial distribution under the root sign in (\ref{n2}) can be approximated by a normal distribution \cite{binom}, and after some algebra (see the "Methods" section (Derivation of Eq.(11))  we have
\begin{eqnarray}\label{9}
A^{\textrm{clock}}_{\textrm{Bose}}(n\gets 0,t)\approx \frac{(-i)^n}{[2\pi n]^{1/4}}\exp\left[-\frac{(t_n-t)^2}{\Delta t^2}\right ],\q
t_n \equiv \om^{-1}\sqrt{n/ N}, \q \Delta t =\om^{-1}N^{-1/2}.
\end{eqnarray}
\begin{figure}[ht]
	\includegraphics[angle=0,width=12cm]{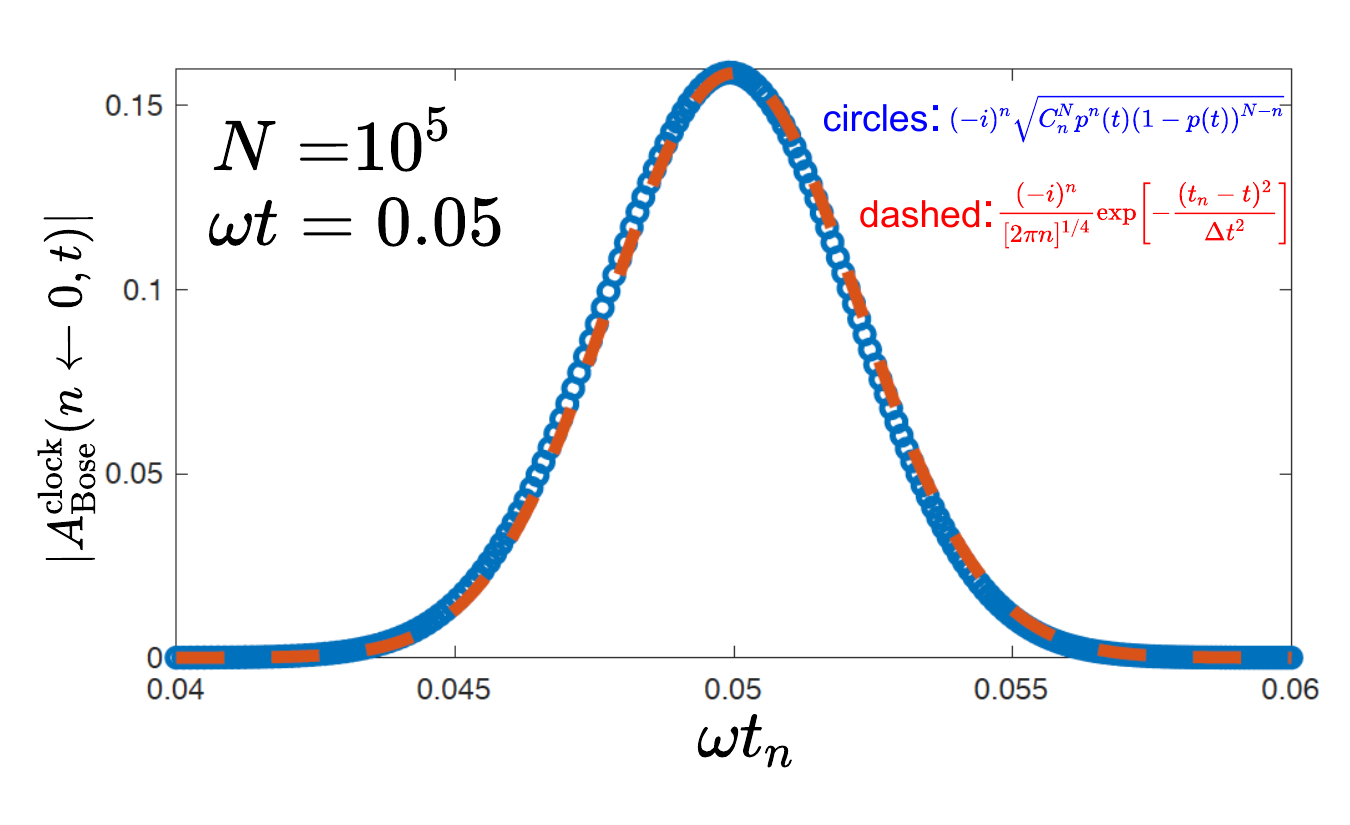}
	\caption{ Comparison between Eqs. (\ref{n2}) and (\ref{9}). \textcolor{black}{Out of $N=10^5$ bosons,  $250$ are in the right well.} }\label{plot:Fig2}
\end{figure}
Alice can now count the atoms in the right well and use $t_n$ in Eq.(\ref{9}) as an estimate for the elapsed time. 
Equation (\ref{9}) shows that her  estimate is likely to be within an error margin $\Delta t$ of the true value $t$.
A good clock is the one which has a small relative error.
If $\omega t$ is kept constant while $N\to \infty$, the error tends to zero, since $\Delta t/t_n =1/\sqrt n \approx 1/(\om t \sqrt{N}) \sim 1/\sqrt{N}$, 
and with many bosons 
Alice has a good clock \textcolor{black}{(see Fig.\ref{plot:Fig2})}. 
\newline
A further remark is in order.
As $N\to \infty$, a large system of independent particles begins to develop certain classical properties \cite{Bruk}, \cite{DSclass}
(see also the "Methods" section (Derivation of Eq.(11)). 
For example, denoting one-partial states
in the wells as $|L\ra$ and  $|R\ra$, and preparing the bosons in a state  $|\Phi^{\textrm{clock}}_{\textrm{Bose}}(0)\ra=\prod_{i=1}^N|L\ra_i$, one later finds them in 
$|\Phi^{\textrm{clock}}_{\textrm{Bose}}(t)\ra=\prod_{i=1}^N[u_{LL}(t)|L\ra_i+u_{RL}(t)|R\ra_i]$, where $u_{LL}(t)=\cos(\om t)$ and $u_{RL}(t)=-i\sin(\om t)$
are the matrix elements of the one-particle evolution operator. The evolved state
$|\Phi^{\textrm{clock}}_{\textrm{Bose}}(t)\ra$ is not an eigenstate of an operator $\hat n =\sum_{i=1}^N |R\ra_i \la R|_i=\hat a^+_R\hat a_R$, which gives the number of bosons in the right well.
However, expanding \textcolor{black}{it} in the eigenstates of \textcolor{black}{$\hat n$, $\hat n |n\ra=n|n\ra$}, $n=0,1...N$, one finds \textcolor{black}{\cite{DSclass}} the coefficients localised in a range $\sim \sqrt{N}$ around a mean value  $\nm(t) \equiv \la \Phi^{\textrm{clock}}_{\textrm{Bose}}(t)|\hat n|\Phi^{\textrm{clock}}_{\textrm{Bose}}(t)\ra= N\sin^{2}(\om t)\approx  N\om^2t^2\textcolor{black}{\propto N}$, 
\begin{eqnarray}\label{9a}
\la n|\Phi^{\textrm{clock}}_{\textrm{Bose}}(t)\ra\approx \frac{(-i)^n}{\left[2\pi \nm(t)\right]^{1/4}}\exp\left[-\frac{\left(n-\nm(t)\right)^2}{2\nm(t)}\right ].
\end{eqnarray}
A similar localisation would occur if $|\Phi(t)\ra$ expanded in any basis, and this has important consequences. 
Firstly, one can accurately measure $\hat n$ (or any other operator \cite{DSclass}) and obtain a result close to its mean  value ($\sim N$) with an error margin $\sim \sqrt N$.
This is a good measurement, since its relative error tends to zero. Secondly, one can measure it  inaccurately, 
e.g., by using a von Neumann pointer prepared in a Gaussian state of a width 
$\sim N^{1/2+\epsilon}$, where $0<\epsilon < 1/2$ \cite{DSclass}. This is still a good measurement since $\sim N^{1/2+\epsilon}/N\to 0$, but also one which in the 
limit $N\to \infty$ leaves the state (\ref{9a}) almost intact, since $N^{1/2+\epsilon}/\sqrt{N}\to \infty$ (see the "Methods" section (A macroscopic clock) for details). Alice can keep reading this macroscopic nearly classical clock 
without affecting its operation, like she would  do with a classical wrist watch.  
\subsection*{A clock which first runs and then stops}
Next Alice needs to make the clock run until the moment the atom emits a photon. 
This can be achieved  by coupling it to the atom-photon Markovian system \textcolor{black}{($M$)} by means of a Hamiltonian
\begin{eqnarray}\label{10}
\h^{\textrm{a+ph+clock}} = \h_{M} + \hat \pi_e \h^{\textrm{clock}}, \q  \hat \pi_e \equiv |e\ra \la e|, 
\end{eqnarray}
where $\hat \pi_e$ projects onto the atom's excited state.  
The corresponding Schr\"odinger equation is easily solved, and the amplitude for the composite  $\{\textrm{(a)tom+(ph)oton + clock}\}$, starting with
the right well empty, to end with $n$
bosons there, 
is found to be (see the "Methods" section (Coupling the clock to a quantum system))
\begin{eqnarray}\label{11}
A^{\textrm{a+ph+clock}}_{\textrm{Bose}}(j,n\gets e,0) = \int_0^t  A^{\textrm{clock}}_{\textrm{Bose}}(n\gets 0,\tau) A^{\textrm{a+ph}}(j\gets e,\textcolor{black}{t|\tau}) d\tau, \n j=e\q\text {or}\q \E,\q n=0,...N,
\end{eqnarray}
where $A^{\textrm{clock}}_{\textrm{Bose}}(n\gets 0,\tau)$ is given by Eq.(\ref{9}), and $A^{\textrm{a+ph}}(j\gets e,\textcolor{black}{t|\tau})$
is the amplitude for the atom-photon system to reach a final state $|e\ra$ or $|\E\ra$ after remaining in $|e\ra$ for exactly $\tau$ seconds,
\begin{eqnarray}\label{12}
A^{\textrm{a+ph}}(j\gets e,\textcolor{black}{t|\tau}) =\la j|\u^{\textrm{a+ph}}(t|\tau)|e\ra,\q\q\q\q\q \n 
\u^{\textrm{a+ph}}(t|\tau)\equiv (2\pi)^{-1}\int_{-\infty}^\infty \exp[i\lm \tau -i(\h_{M}+\lm \hat \pi_e)t ] d\lm.
\end{eqnarray}
where $\u^{\textrm{a+ph}}(t|\tau)$ is the conditional evolution operator. This is clearly the desired result. The clock runs only while the atom remains in the excited state, and the amplitudes are added for all possible 
durations $\tau$, which may lie between $0$ and $t$. The integral in Eqs.(\ref{12}) is evaluated  by noting that adding 
$\lm \hat \pi_e$ to $\h_{M}$ only shifts the energy of the discrete state $E_e$ by $\lambda$ (see the "Methods" section (Timing the transition in the Markovian case)). The result 
($0\le \tau \le t$)
\begin{eqnarray}\label{13}
A^{\textrm{a+ph}}(e\gets e,t|\tau) =\exp(-iE_e t -\G t/2)\delta(\tau-t),\q\q\q\q\q\q\q\n 
A^{\textrm{a+ph}}(\E\gets e,t|\tau) =-i\Om \exp[-i\E (t -\tau)] \exp[-i (E_e-i\G /2) \tau]
\end{eqnarray}
confirms what is already known from  Eqs.(\ref{2c}) and (\ref{3c}).
An atom, still found  in the excited state at $t$, has  remained in that state all the time. An atom, found in the ground state, has not  returned to the excited state after making a single transition at some $\tau$ between $t=0$ and $t$. 
\newline
Alice  the practitioner can now prepare the atom in its excited state, couple 
\textcolor{black}{it with} a \e{good} clock (\ref{9}), wait until time $t$, and then measure the  energy of the photon (if any), as well as count the bosons in the right well.
She can find no photon and $n$ bosons, with a probability
\begin{eqnarray}\label{14}
P(e,n\gets e,0) =\exp(-\G t)P^{\textrm{clock}}_{\textrm{Bose}}(n\gets 0,t),  \q \sum_{n=0}^N P_{\textrm{Bose}}(n\gets \textcolor{black}{0},t)=1,
\end{eqnarray}
where $P^{\textrm{clock}}_{\textrm{Bose}}(n\gets 0,t)=|A^{\textrm{clock}}_{\textrm{Bose}}(n\gets 0,t)|^2$ (see Eq.(\ref{9})).
She may find $n$ bosons,  a photon with an energy $\E$,  and conclude that the emission occurred 
around \textcolor{black}{(see Eq. (\ref{9}))}
\begin{eqnarray}\label{14a}
\tau _n =\om^{-1}\sqrt{n/N}.
\end{eqnarray}
The error of this result is determined by the width of the Gaussian (\ref{9}) which,  restricts the possible values of $\tau$ in Eq.(\ref{11}). Alice's relative error is, therefore, $\Delta t /\tau_n \sim 1/\sqrt{n}<<1$, where
$\Delta t =\om^{-1}N^{-1/2}$ was defined in Eq.(\ref{9}).
The probability of this outcome is given by the absolute square of $A^{\textrm{a+ph+clock}}_{\textrm{Bose}}(\E,n\gets e,0)$ in Eq.(\ref{11}).
Extending in Eq.(\ref{11}) the limits of integration to $\pm \infty$, and evaluating Gaussian integrals yields
\begin{eqnarray}\label{15}
P(\E,\tau_n\gets e,0)\approx \frac{\pi \Om^2}{\om \sqrt{nN}}\exp(-\G \tau_n)\times \frac{\Delta t}{\sqrt{2\pi}}\exp[-(\E-E_\d)^2\Delta t^2/2]
\end{eqnarray}
for $0<\tau_n < t$, and $P(\E,\tau_n\gets e,0)=0$ otherwise. 
\newline
The net probability of an outcome $\t_n$ is
\begin{eqnarray}\label{16}
P(\tau_n\gets e,0)= \int d\E \rh(\E) P(\E,\tau_n\gets e,0) \approx 
\frac{ \G}{2\om \sqrt{nN}}\exp[-\G \tau_n]
\end{eqnarray}
and replacing $\sum_n \to \int_0^t d\tau_n$
helps to verify that the overall decay rate is not affected by the presence of the clock,
$P^{\textrm{decay}}_{\Delta t}(t)=\sum_n P(\tau_n\gets e,0)=1-\exp(-\G t)$. Finally, the spread of the energies of the emitted photons
is no longer Lorentzian, but Gaussian,
\begin{eqnarray}\label{16a}
P(\E\gets \d, t\to \infty)= \sum_n P(\E,\tau_n\gets e,0) \approx \frac{\Delta t}{\sqrt{2\pi}}\exp[-(\E-E_\d)^2\Delta t^2/2],
\end{eqnarray}
and becomes broader as Alice's accuracy improves, $\Delta t \to 0$. 
[Note that we cannot arrive at the Lorentzian distribution (\ref{5c}) simply by sending $\Delta t \to \infty$ in Eq.(\ref{16a}), 
since Eq.(\ref{16}) was derived under assumption that the number of bosons in the right well is large.]
\subsection*{A clock which first waits and then runs } \label{clockWaits}
Alice can also consider a Markovian clock which starts running only after the transition has taken place and continues doing so 
\textcolor{black}{until} 
the time of observation $t$. (It will be clear \textcolor{black}{shortly} why this 
case \textcolor{black} {is} 
of interest
). Replacing in Eq.(\ref{10}) projector $\hat \pi_e$ by $1-\hat \pi_e=\int_{-\infty}^{\infty} dE_r |E_r \ra\la E_r |$,
$\tau$ with $t-\tau$, and
acting as before
yields (see the "Methods" section (Coupling the clock to a quantum system)) 
\begin{eqnarray}\label{1w}
A^{\textrm{a+ph+clock}}_{\textrm{Bose}}(e,n\gets e,0) = \exp[-i E_et-\G t /2] \delta_{n0},\q\q\q\q\q\q\q\q\q\q\n
A^{\textrm{a+ph+clock}}_{\textrm{Bose}}(\E,n\gets e,0) \approx
\frac{(-i)^{n+1}\Om}{[2\pi n]^{1/4}}\times \q\q\q\q\q,\q\q\q\q\q\q\q\q\q\n
\int_0^t \exp\left[-\frac{(t_n-\t)^2}{\Delta t^2}\right ] \exp\{-i\E\t -i(E_e-i\G /2)(t-\tau)]\}d\tau,\q\q\q\q\q
\end{eqnarray}
where $\delta_{n0}$ is the Kronecker delta.
Now the number of the bosons in the right well is determined by the time which has elapsed since the moment of emission, 
and we can attend to the cat which dies as a result of the atom's decay.
\newline
\subsection*{Exploding powder kegs and poisoned cats} 
\begin{figure}[ht]
	\includegraphics[angle=0,width=14cm]{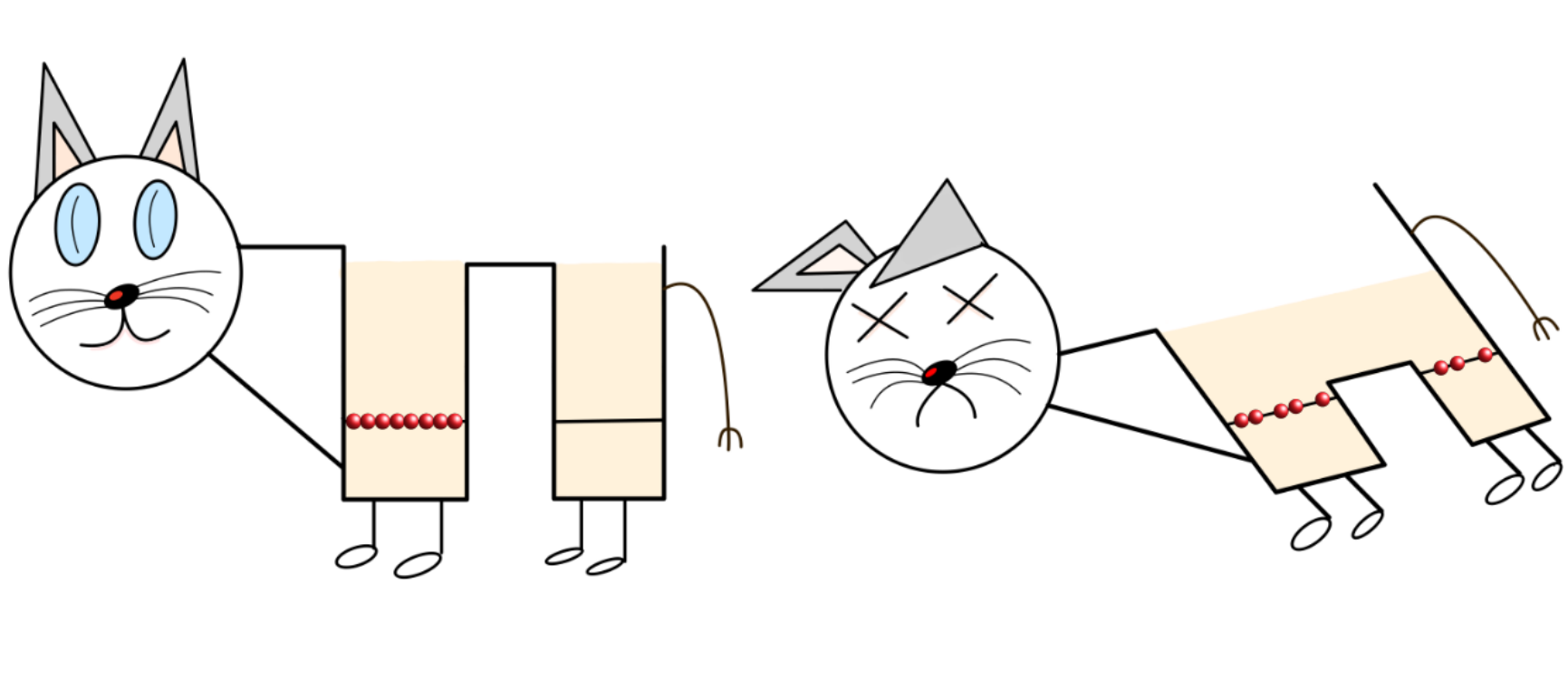}
	\caption{ An artists's impression of a primitive cat a) alive and well, and b) sadly, dead for some time. 
		Any resemblance to real cats, living or dead, is purely coincidental.    
	} \label{plot:Fig4a}
\end{figure}

It is difficult to resist the temptation to relate the present discussion to the famous Schr{\"o}dinger's Cat problem.
In 1935 Einstein and Schr{\"o}dinger discussed a hypothetical case in which explosion of a powered keg was caused 
by a photon emitted by a decaying atom. In \cite{Cat} Schr{\"o}dinger dramatised the narrative further by replacing 
the unstable powder by a now famous live cat, which dies in the event. The perceived contradiction 
was due to the fact \textcolor{black}{that,} 
prior to the final observation of the cat's state\textcolor{black}{,} the wave function of the joint system 
was deemed to be a superposition of the states $|\text{atom: excited}\ra\otimes |\text{cat: alive}\ra$ and $|\text{atom: decayed}\ra\otimes |\text{cat: dead}\ra$. 
With wave function  believed to reflect on the actual condition of a system,
this left  a big question mark over the cat's situation prior to be found either dead or alive. The same contradiction was observed in the powder keg example, 
where, again, macroscopically distinguishable states $|\text{unexploded}\ra$ and $|\text{exploded}\ra$ were forced into superposition through  entanglement with the atom. 
\newline 
It is worth revisiting the situation by replacing the cat (the keg) with the (nearly) classical clock of Section \ref{clockWaits}. 
So far, the cat paradox did not arise because we only required
a matrix element of a unitary operator 
$\u^{\textrm{a+ph+clock}}(t) =\exp(-i \h^{\textrm{a+ph+clock}}t)$ between the states $|\E\ra \otimes |n\ra$  and $|e\ra \otimes |0\ra$ in the Hilbert space of the composite ${\textrm{a+ph+clock}}$. 
The question, we recall, was 
\e{is there a photon, and how many bosons are there in the right well at $t$?} 
Although there appears to be no need for it, one can create a kind of \e{cat} problem by looking at the ket
\begin{eqnarray}\label{1u}
\u^{\textrm{a+ph+clock}}(t)|e,0\ra=   \exp[-i E_et-\G /2t] |\Phi^{\textrm{clock}}_{\textrm{Bose}}(0)\ra\otimes |e\ra+	\n
\sum_r \int_0^t d\tau' A^{\textrm{a+ph}}(\E\gets e,t|t-\tau')  |\Phi^{\textrm{clock}}_{\textrm{Bose}}(\t)\ra \otimes |\E\ra
\end{eqnarray}
and \textcolor{black}{object} the appearance of a superposition of distinguishable macroscopic states in r.h.s. of Eq.(\ref{1u}).  Indeed,  
for an accurate clock, i.e. $\Delta t \to 0$ ($N>>1$), the clock's states in the r.h.s. of Eq.(\ref{1u}) are practically orthogonal [cf. Eq.(\ref{9a})], 
$\la \Phi^{\textrm{clock}}_{\textrm{Bose}}(\tau')|\Phi^{\textrm{clock}}_{\textrm{Bose}}(\t)\ra \sim 
\exp\left [-{(\t-\t')^2}/{\Delta t^2}\right]\xrightarrow[\Delta t \to 0]{} 0$.
Alternatively, one can avoid the paradox of the cat being both dead and alive by considering
the superposition to be a transient artefact of the calculation, needed only
to establish the likelihood of finding $n$ escaped bosons, and having no further significance.  
\newline
The analogy can be taken further. Neither the cat's demise, nor an explosion are purely instantaneous events.
By looking at the deterioration of the cat's body (we leave outside the question of what it means to be alive)
one can tell how long ago it stopped functioning. By looking at how much of the powder has been burnt, or how much dust thrown up in the air 
has settled, 
it is possible to deduce the moment when explosion started. 
Remarkably, the waiting clock of \textcolor{black}
Section \ref{clockWaits} keeps a similar record, only in a more direct way 
\textcolor{black}{(Fig.\ref{plot:Fig4a})}. 
Alice may find no bosons in the right well (cat is alive), or a certain number of them (a particular  stage
of decay of the dead cat's body). The more accurately Alice is able to deduce the \e{moment of death}, 
the broader will be the energy distribution of the photon whose emission  has killed the cat [cf. Eq.(\ref{16a})]. 
A valid analogy could be  a very long fuse, whose  burnt length (number of bosons in the right well)  
would let one deduce the moment when it was set on fire.

\subsection*{Beyond the wide band  approximation}
\begin{figure}[ht]
	\includegraphics[angle=0,width=14cm]{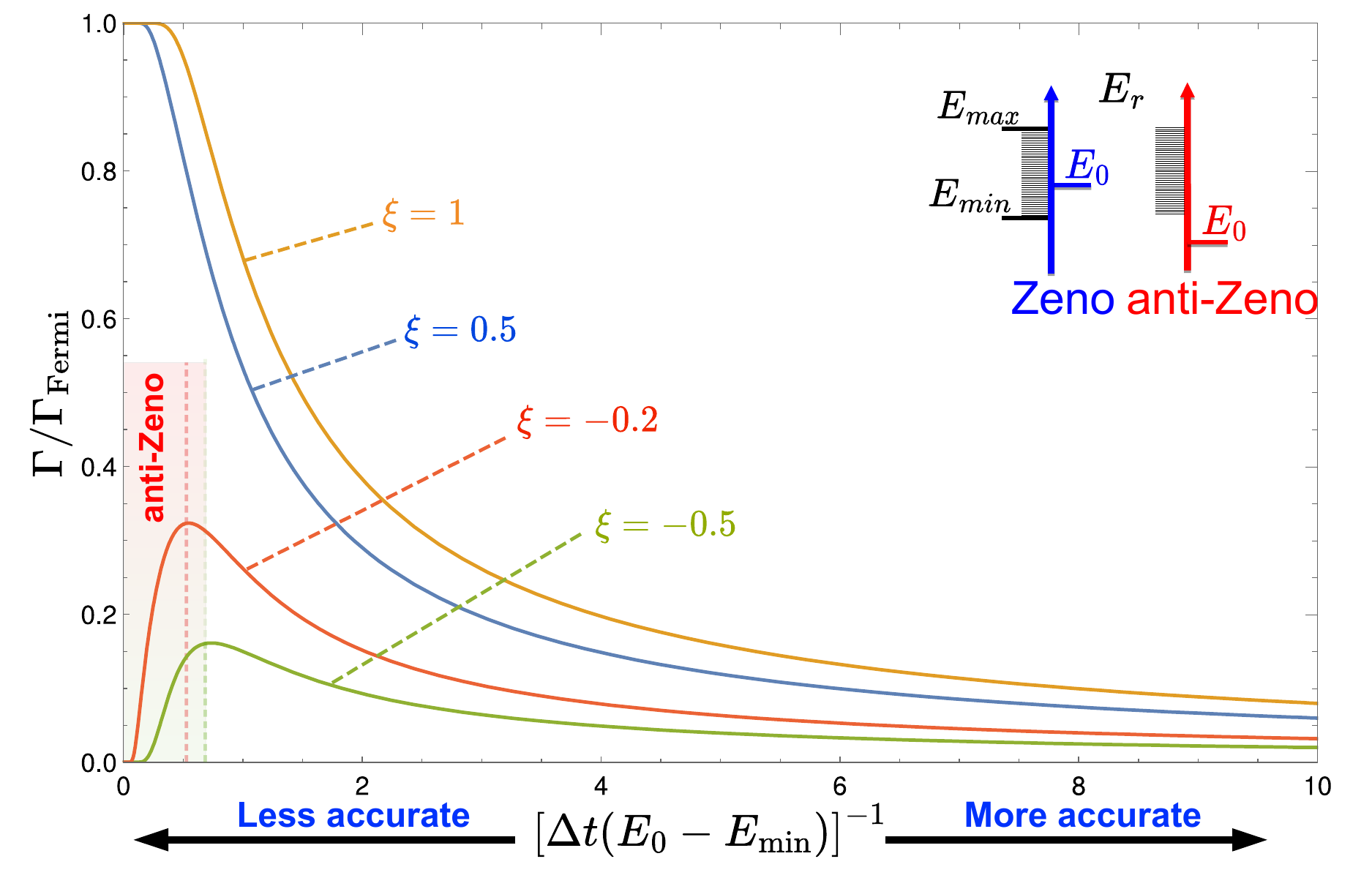}
	\caption{ The rate of decay into a finite-sized  band $E_{\textrm{min}}\le E_r\le E_{\textrm{max}}$ as a function of the clock's accuracy $\Delta t$
		[$\xi = 2(E_e-E_{\textrm{min}})/\Delta E_r$]. For $E_{\textrm{min}}< E_e< E_{\textrm{max}}$ better accuracy means a smaller decay rate (Zeno effect); 
		for  $E_e < E_{\textrm{min}}$ there is an initial increase in the value of $\Gamma$ (anti-Zeno effect). Both possibilities are illustrated in the inset. \textcolor{black} {The anti-Zeno effect also occurs in the case $E_e >E_{\textrm{max}}$, not shown here. }}
	\label{plot:Fig4}
\end{figure}

Next we revisit 
a more general (non-Markovian) case of Section II [cf. Eq.(\ref{2b})],
where the product $|\Om(\E)|^2\rh(\E)$ may depend on the photon's energy, 
$\int_{-\infty}^\infty  dE |\Om(\E)|^2\rh(\E)$ is finite, and the transition occurs via a single jump.
Only a small proportion of all atoms will be found
decayed by the time $t$, but Alice may still want to know when this unlikely transition did occur. 
A simple calculation (see the "Methods" section (Timing the first order transition in a non-Markovian case))
shows that the probability of the clock's reading $\t_n$  for a  system ending in a state $|\E\ra$, is still 
given by an expression similar to Eq.(\ref{15}),   
\begin{eqnarray}\label{1e}
P(\E,n\gets e,0,t)\approx \frac{\pi \Om^2(\E) \Delta t^2}{[2\pi n]^{1/2}} \exp[-(\E-E_e)^2\Delta t^2/2],
\end{eqnarray}
so that measuring the moment of emission to an accuracy $\Delta t$ broadens the range of the photon's energies, 
which grows as $1/\Delta t$ owing to the Gaussian in r.h.s. of Eq.(\ref{1e}). Therefore, it is the availability of the final system's states that restricts the decay rate, 
and is responsible for the Zeno effect already mentioned in Sect.II. Indeed, 
acting as before \textcolor{black}{(cf. Section V)}, 
for the probability to decay by the time $t$ we find
\begin{eqnarray}\label{2e}
P^{\textrm{decay}}_{\Delta t}(t) =\sum_n\int d\E \rh(\E)P(\E,n\gets e,0,t)=  \G_{\Delta t}\times t, 
\end{eqnarray}
\begin{eqnarray}\label{2ea}
\G_{\Delta t} \equiv \sqrt{2\pi} \Delta t \int
d\E \rh(\E)\Om^2(\E) \exp[-(\E-E_\d)^2\Delta t^2/2].
\end{eqnarray}
In the Markovian wide band limit $\G_{\Delta t}$ in Eq.(\ref{2ea}) does reduce to Fermi's golden rule \cite{Fermi}, $\G_{\Delta t}=\G_{\textrm{Fermi}}=2\pi  \Om^2 \rh$.
But if the integration of an ever broader Gaussian is restricted to a finite range, the factor of  $\Delta t$ in Eq.(\ref{2ea})
is no longer cancelled, and 
the decay rate eventually decreases  as the measurement becomes
more accurate. 
For example, consider a special case of an energy band of a width $\Delta \E
=E_{\textrm{max}}-E_{\textrm{min}}$, 
wherein  $\rh(\E)\Om^2(\E)=const$.
Comparing the decay rates prescribed by Eq.(\ref{2ea}) and by Fermi's rule, we have 
\begin{eqnarray}\label{4e}
\G_{\Delta t}/\G_{\textrm{Fermi}}\xrightarrow[\Delta t \to 0]{}\frac{\Delta \E\Delta t}{\sqrt{2\pi}}.
\end{eqnarray}
The accuracy with which the moment of emission can be determined without significantly altering the decay rate is, ultimately, 
limited  by the width of the energy range, available to the emitted photon. What happens for not too small values of $\Delta t$
depends, however, on whether the excited atom's energy lies within the allowed range, as explained in Fig.\ref{plot:Fig4}. 
If $E_e <E_{\textrm{min}}$ or \textcolor{black}{$E_e >E_{\textrm{max}}$}
unobserved atom cannot decay, and the decay rate first increases as $\Delta t$ becomes smaller, 
leading to a kind of \e{anti-Zeno} effect \cite{Anti}. 
It  eventually begins to \textcolor{black}{fall off in} agreement with Eq.(\ref{4e}),
when the exponential in Eq.(\ref{1e}) can be approximated by unity. 
\newline
With all this in mind, we can revisit the analysis of \cite{Schul}, where the {\it duration} of a jump was estimated in the following manner.
Every $\delta t$ seconds one checks whether the atom continues in its excited state. The jump time, $\tau_{\textrm{J}}$, is then taken to 
be the $\delta t$ for which the checks begin to affect the atom's decay rate. For  $\tau_{\textrm{J}}$, \cite{Schul} finds
\begin{eqnarray}\label{6e}
\tau_{\textrm{J}}\approx \G_{\textrm{Fermi}}\t_{\textrm{z}}^2,  
\end{eqnarray}
where $\t_{\textrm{z}}\equiv \left [\la e|\h^2_{\textrm{a+ph}}|e\ra -\la e|\h_{\textrm{a+ph}}|e\ra^2\right ]^{-1/2}$ is the \e{Zeno time}.
In the regime studied in \cite{Schul} $\delta t$ is short enough for the transition to occur via a single jump.
One can, therefore, equally interpret  $\tau_{\textrm{J}}$ as  the {\it uncertainty} in the moment at which an instantaneous transition takes place.
In the case we have studied here, Alice's clock begins to affect the decay rate when its error  
is of order of the inverse band width, $\Delta t \sim 1/\Delta E_r$ [cf. Eq.(\ref{4e})].
It is easy to check (see the "Methods" section (The "jump time" in Eq.(28))) that  Eq.(\ref{6e}) yields a similar  result,
\textcolor{black}{$\tau_{\textrm{J}} \sim 1/\Delta E_r $}.
\subsection*{Feynman's \e{only mystery of quantum mechanics}}  
\begin{figure}[ht]
	\includegraphics[angle=0,width=13cm]{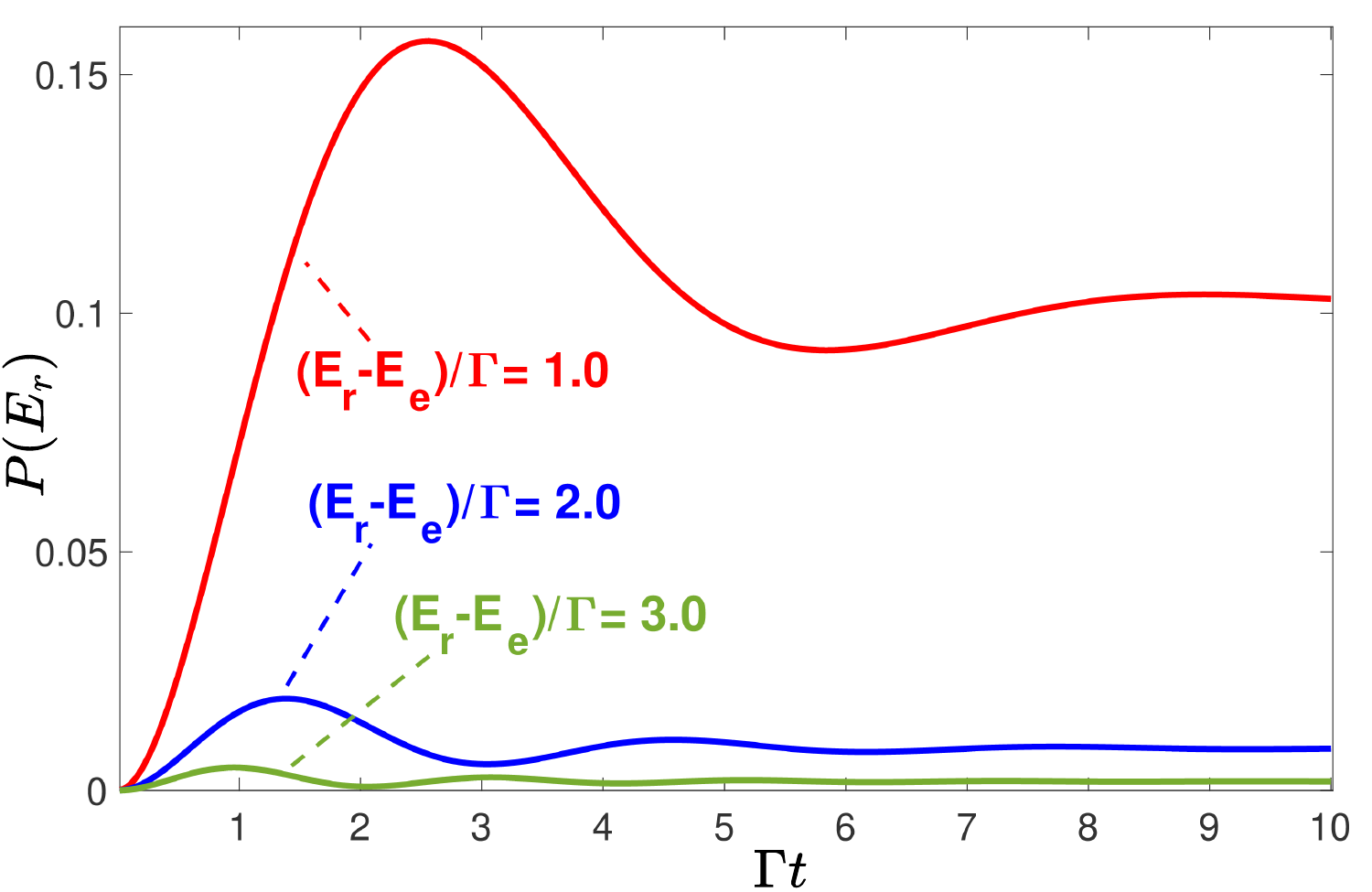}
	\caption{ Probability of finding the photon in a unit interval around energy $\E$ in a single measurement made at time $t$.}\label{plot:Fig3}
\end{figure}

\textcolor{black}{All this  leaves one with a question: \e{what can be said about the moment of emission if it has not been timed 
		by a cat, gunpowder, or a clock?
}}
Very little, according to Feynman \cite{FeynL}, \cite{FeynC}. In a double slit experiment a particle can reach a point on the screen by passing through the holes, with the probability amplitudes $A_1$ and $A_2$, respectively. The probability of arriving \textcolor{black}{at} the screen with both slits open is $|A_1+A_2|^2$, while with only the first one open it is $|A_1|^2$. With no restriction on the signs of the amplitudes, it is possible to have (e.g., near a dark fringe)
$|A_1+A_2|^2<|A_1|^2$, so that eliminating one of the routes increases the number of arriving particles. For this reason, it is not possible
to assume that a setting of the particle's internal machinery (or any other hidden variable) predetermines the hole to be chosen by each particle
on its way to the screen. The mathematics cannot be simpler, and one must conclude that \e{... when you have no apparatus to determine through which hole the thing goes, then you cannot say that it either
	goes through one hole or the other}. This is an illustration of the Uncertainty Principle \cite{FeynL} which states 
that one cannot determine which of the alternatives has been taken without destroying interference between them.
\newline
The same principle, applied to the case of a decaying atom, states that with no apparatus to determine the moment of decay, one cannot say that the atom emits a photon with an energy between $\E$ and  $\E+d\E$ at one moment or the other. Indeed, if each atom were predestined to decay at a given time, 
the number of decayed atoms could only {\it increase} or stay the same as
the time span available for the atom's decay becomes longer. 
However, the corresponding probability is given by $W(\E,\E+d\E)=P(\E) d\E$, and $P(\E)=\rho |\int_0^t A^{\textrm{a+ph}}(\E\gets e,t|\t ) d\t|^2$,  shown in 
Fig.\ref{plot:Fig3}, can {\it decrease} with $t$.
(Note that the probability in Fig.\ref{plot:Fig3} is that of a {single} measurement made at different times.
If the decayed atoms are counted twice, the number measured at a later time is, of course, always greater.)
The decrease cannot be blamed on 
the re-absorption of the photon, impossible in the Markovian model [cf. Eq.(\ref{2ca})]. Neither can it be explained by the change in the emitted photon's energy [cf. Eq.(\ref{2cb})].
This seems even stranger than the double slit case. One could imagine the routes passing through different holes 
merged, like two confluent rivers, where it is impossible to say on which of the two a boat is.
Merging time intervals may be even more difficult to fathom, but to conclude that an unobserved  transition has occurred at a particular moment would lead to \e{an error in prediction}\cite{FeynC}, as was discussed above.
This is, according to Feynman \cite{FeynL}, the only mystery of quantum mechanics, 
which defies \e{any deeper explanation}.
\section*{Conclusions}
\noindent
The story of the Schr\"odinger's cat, whose death is caused by the decay of an excited atom, is one of the best known illustration
of a problem which one expects to arise when the classical world meets its quantum counterpart \cite{Cat}. 
A classical system, believed to have an unbroken continuous history, appears to loose this property if forced to interact with a quantum
object, for which no continuous description is thought to be available \cite{Copen}.
To bridge the gap between the classical and quantum views we design a nearly classical macroscopic clock, capable of timing the 
moment of decay to a good, yet finite accuracy. The complete narrative is as follows. 

An atom, prepared in its excited \textcolor{black}{state} is found decayed at time $t$, after having emitted a photon with energy $E_r$.
The instant of emission is unknown, and to determine it the experimenter needs a device which would measure it. 
One suitable choice is a clock consisting of large number $N$ of noninteracting bosonic atoms, initially trapped in the left 
well of a double well potential. Finding that  $n$ bosons have made the transition to the right well, one can estimate the
elapsed time $t$ as $t_n  \approx \om^{-1}\sqrt{n/N}$, with an error $\Delta t  \approx \om^{-1}/\sqrt{N}$.
With the transition amplitude $\om$ small, and  the number of bosons large, $N>>1$  \textcolor{black}{the clock is a source of irreversible current flowing from left to right.} With many bosons in the right well, $N>>n>>1$, the clock is seen to acquire an important  classical property. Its wave function becomes localised,  
and one is able to measure time to a good accuracy without significantly perturbing the clock's evolution (for more details see also \cite{DSclass}). 
\newline 
The clock can be arranged to run until the moment of emission,
which  would yield a good estimate of the time of emission provided $\Delta t/t_n << 1$,  except in the unlikely case of the decay occurring almost immediately. 
The effect of the measurement on the atom's decay depends on the range of energies $\Delta E_r$, available to the emitted  photon.
In the wide band limit, $\Delta E_r\Delta t >>1$ the decay rate $\Gamma$ remains the same, and destruction of interference between 
the moments of emission leads only to  broadening of the photon's energy spectrum, whose shape is no longer Lorentzian, but 
Gaussian,  with a width $\sim 1/\Delta t$. 
Having obtained a result $t_n$, and  knowing that more measurements could have been added both before and after $t=t_n$
(almost) without altering the clock's evolution, the experimenter has a complete history of what has happened. 
The atom remained in its excited state until $t_n-\Delta t \lesssim  t_n+\Delta t$, and then continued in the ground state
until the time when the  clock is read. Note that essential for recovering such a continuous description is the classical property of the 
macroscopic clock reached in the limit $N>>1$. 
\newline
The Zeno effect sets in when the inverse clock's accuracy become comparable to the range of available photon's energies, $\Delta E_r\Delta t\lesssim1$.
Now the notion of the moment of decay is meaningful only in the weak coupling limit,  $\Gamma_{\textrm{Fermi}} t <<1$ [cf. Eq.(\ref{2b})].
In the \e{narrow band} limit, $\Delta E_r\Delta t << 1$,  the decay rate is proportional to $\Delta t$, and the unlikely atomic decay is further suppressed as $\Delta t \to 0$.
\newline
The clock  set up to run {after} the decay has occurred, helps provide an additional insight into  the fate of the Schr\"odinger's feline \cite{Cat}. 
Now one knows that  there were no bosons in the right well until $t_n$ (within an error margin $\Delta t$), 
after which their number there was steadily growing. One can leave the question of what it means to be alive outside the scope 
of quantum theory, and concentrate instead on the deterioration of the cat's macroscopic physical body.
The  waiting clock is a blueprint for a very primitive \e{cat}, said to be alive if there are no bosons in the right well, $n=0$, and dead in some stage of decay 
with $n>>1$. 
If the analogy holds, a real cat's physical frame should be characterised by a quantum uncertainty $\Delta t_{\textrm{cat}}$, 
which limits the ability of an experienced forensic scientist to determine the time of death by studying the cat's remains. 
The cat's fate depends, therefore, on the details of the atom's decay. 
In the wide band model, the probability of survival up to time $t$ decreases as $\exp(-\Gamma t)$, regardless of the $\Delta t_{\textrm{cat}}$.
However, in the finite band case, 
a cat whose body would allow to determine the moment of death 
with greater precision should have a better chance to survive its ordeal.

\section*{Methods}

\subsection*{
	Derivation of Eq.(\ref{9})}
The normal approximation to $A^{\textrm{clock}}_{\textrm{Bose}} (n\gets 0,t)= (-i)^n\sqrt {C^N_n p^n(1-p)^{N-n}}$ reads ($p(t)=\om^2t^2<<1$)
\begin{eqnarray}\label{A1}
A^{\textrm{clock}}_{\textrm{Bose}}(n\gets 0,t)\approx(-i)^n(2\pi Np)^{-1/4}\exp[-(n-Np)^2/4Np].
\end{eqnarray}
In new variables $t_n\equiv \om^{-1}\sqrt{n/N}$ and $\Delta t \equiv  \om^{-1}N^{-1/2}$ we have
\begin{eqnarray}\label{A2}
A^{\textrm{clock}}_{\textrm{Bose}}(n\gets 0,t)\approx(-i)^n (2\pi N \om^2t^2)^{-1/4}\exp[-(t^2_n-t^2)^2/4t^2\Delta t^2].
\end{eqnarray}
As $N\to \infty$, $\Delta t \to 0$, and the exponential is  sharply peaked around $t_n \sim t$, or $n\sim N \om^2 t^2$, 
and the amplitude can be approximated by
\begin{eqnarray}\label{A3}
A^{\textrm{clock}}_{\textrm{Bose}}(n\gets 0,t)\approx(-i)^n(2\pi n)^{-1/4}\exp[-(t_n-t)^2/\Delta t^2],
\end{eqnarray}
which is Eq.(\ref{9}).
\subsection*{
	A macroscopic clock}
Consider $N$ non-interacting bosons, each occupying the same state $|\phi\ra$,  $|\Phi\ra=\prod_{i=1}^N |\phi\ra_i$.
Expanding the $|\phi\ra$s in an orthonormal basis $|j\ra$, $j=1,2$, $|\phi\ra= \alpha |1\ra +\beta|2\ra$ yields
\begin{eqnarray}\label{B1}
|\Phi\ra= \sum_{\nn=0}^N B_{\nn}|\nn,N\ra,\q B_{\nn}=\sqrt{C^N_{\nn}} \alpha^{\nn}\beta^{N-\nn},
\end{eqnarray}
where $C^N_{\nn}$ is the binomial coefficient, and $|\nn,N\ra$ describes a state with $\nn$ particles populating the state $|1\ra$.
Suppose one wants to measure the number of particles in the state $|1\ra$, 
$\hat N_{1} =\sum_{i=1}^N |1\ra_i\la 1|_i$, 
using a Gaussian von Neumann pointer, whose initial state is
$G(f)=C \exp[-f^2/\Delta f^2]$. After the measurement for the entangled state of the pointer and the bosons, $|\Phi\ra$, one finds
\begin{eqnarray}\label{B2}
\la f|\Phi\ra= \sum_{\nn=0}^N B_{\nn}G(f-\nn)|\nn,N\ra.
\end{eqnarray}
The distribution of the pointer's readings $f$ and the mixed state of the bosons $\hat\rho$ are, therefore, given by 
\begin{eqnarray}\label{B3}
w(f)= \sum_{\nn=0}^N |B_{\nn}|^2G^2(f-\nn),
\end{eqnarray}
and 
\begin{eqnarray}\label{B4}
\hat\rho =\sum_{\nn, \nn'=0}^N B^*_{\nn'}B_{\nn}I_{\nn'\nn}|\nn,N\ra\la \nn',N|, \q I_{\nn'\nn}\equiv\int G(f-\nn')G(f-\nn)df.
\end{eqnarray}

With $N,|\alpha|^2N >>1$ the readings lie near the mean value $ \nnn=|\alpha|^2N$.
Using the normal approximation for the binomial distribution $|B_{\nn}|^2$, and replacing the sum by an integral yields
\begin{eqnarray}\label{B5}
w(f)\approx\frac{C^2}{\sigma\sqrt{2\pi}} \int  \exp\left[ -\frac{(\nn-\nnn)^2}{2\sigma^2}-\frac{2(f-\nn)^2}{\Delta f ^2}\right ]d\nn
\sim \exp\left [ -\frac{2(f-\nnn)^2}{\Delta f^2+4\sigma^2}\right ],
\end{eqnarray}
where $\sigma=\sqrt {N |\alpha|^2(1- |\alpha|^2)}$.  
For a large $N$ it is possible to choose $\sigma << \Delta f << \nnn$. 
This yields a good measurement, $w(f)\sim \exp\left [-2(f-\nnn)^2/\Delta f^2\right ]$, with a relative error $\sim \Delta f /\nnn <<1$. 
What is more, since the non-zero $B_{\nn}$s lie within a range $\sim \sigma$ around $\nnn$, all relevant factors 
$I_{\nn'\nn}$ in Eq.(\ref{B4}) can be replaced by unity. Thus, the bosons' state is almost unperturbed by a good, 
yet weakly perturbing measurement, and is ready for the next observation. Since the choice of the basis $|j\ra$ is arbitrary, 
one can say that, for a large system, different collective (macroscopic) variables acquire well define \e{classical} values even when 
the corresponding one-particle projectors $\hat \nn=|1\ra\la1|$ and $\hat n_{1'}=|1'\ra\la1'|$ do not commute. 
By the same token, the progress of a large system can be monitored by consecutive measurements of the same macroscopic quantity without 
seriously affecting its evolution. This is \e{classicality by numbers} \cite{DSclass}.
\subsection*{
	Coupling the clock to a quantum system}
Consider an evolution operator for  a system (s) coupled to a clock, 
\begin{eqnarray}\label{C1}
\u^{\textrm{s+clock}}(t)=\exp[-i(\h^{\textrm{s}} + \hat \pi \h^{\textrm{clock}})t], 
\end{eqnarray}
where $ \hat\pi$ projects onto a sub-space $h$ of the system's Hilbert space.  Since $\h^{\textrm{clock}}$ commutes with both 
$\h^{\textrm{s}}$ and $\hat \pi$ we can write ($\delta(x)$ is the Dirac delta)
\begin{eqnarray}\label{C2}
\u^{\textrm{s+clock}}(t)=\int_{-\infty}^\infty d\lm \delta(\lm -  \h^{\textrm{clock}})  \exp[-i(\h^{\textrm{s}} +\lambda \hat \pi)t].
\end{eqnarray}
But 
$\delta(\lm -  \h^{\textrm{clock}})=(2\pi)^{-1}\int_{-\infty}^\infty d\t\exp(i\lm\t- \h^{\textrm{clock}}\t)$, 
and we have 
\begin{eqnarray}\label{C3}
\u^{\textrm{s+clock}}(t)=\int_{-\infty}^\infty d\t \u^{\textrm{s}}(t|\t)\u^{\textrm{clock}}(\t),
\end{eqnarray}
where $ \u^{\textrm{s}}(t|\t)=(2\pi)^{-1}\int_{-\infty}^\infty d\lm \exp(i\lm\t) \exp[-i(\h^{\textrm{s}} + \lm \hat \pi)t]$
evolves the system under an additional condition that it must spend $\t$ seconds in the chosen sub-space, 
and $\u^{\textrm{clock}}(\t)=\exp(-i \h^{\textrm{clock}}\tau)$ evolves the clock for precisely $\t$ seconds. 
\newline 
If the clock is set to measure the duration spent by the system in the sub-space orthogonal to $h$,
$ \hat \pi$ is replaced by $1-\hat \pi$, and Eq.(\ref{C3}) becomes
\begin{eqnarray}\label{C4}
\u^{\textrm{s+clock}}(t)=\int \u^{\textrm{s}}(t|\t)\u^{\textrm{clock}}(t-\t)d\t = \int  \u^{\textrm{s}}(t|t-\t)\u^{\textrm{clock}}(\t)d\t 
\end{eqnarray}
with the clock running whenever the system is {\it not } in the subspace $h$.
For a transition amplitude between states $|\psi^{\textrm{s}}_i\ra|\phi^{\textrm{clock}}_i\ra$ and $|\psi^{\textrm{s}}_f\ra|\phi^{\textrm{clock}}_f\ra$,
$A^{\textrm{s+clock}}(\psi^{\textrm{s}}_f,\phi^{\textrm{clock}}_f\gets \psi^{\textrm{s}}_i,\phi^{\textrm{clock}}_i,t)= \la \psi^{\textrm{s}}_f|\la\phi^{\textrm{clock}}_f|\exp(-i\h^{\textrm{s+clock}}t)|\psi^{\textrm{s}}_i\ra|\phi^{\textrm{clock}}_i\ra$
we have 
\begin{eqnarray}\label{C5}
A^{\textrm{s+clock}}(\psi^{\textrm{s}}_f,\phi^{\textrm{clock}}_f\gets \psi^{\textrm{s}}_i,\phi^{\textrm{clock}}_i,t)=\int  A^{\textrm{s}}(\psi^{\textrm{s}}_f\gets \psi^{\textrm{s}}_i,t|\t)A^{\textrm{clock}}(\phi^{\textrm{clock}}_f\gets \phi^{\textrm{clock}}_i,\t)d\tau,
\end{eqnarray}
where $A^{\textrm{s}}(\psi^{\textrm{s}}_f\gets \psi^{\textrm{s}}_i,t|\t)$ is the amplitude of the system found in its final state while spending a duration $\tau$ in $h$,
and $A^{\textrm{clock}}(\phi^{\textrm{clock}}_f\gets \phi^{\textrm{clock}}_i,\t)$ is that of the clock reaching $|\psi^{\textrm{s}}_i\ra$ after $\tau$ seconds. 
For a clock measuring the duration spent in the part of the Hilbert space, orthogonal to $h$, $\tau$ should be replaced by 
$t-\tau$ as it has been done in \textcolor{black}{Eq.(\ref{C4})}.
\subsection*{
	Timing the transition in the Markovian case}
Now the system including the atom and a photon (if any) is described by the Hamiltonian (\ref{1b}), $\hat \pi_e \equiv |e\ra \la e|$
projects onto the atom's excited state (no photons). Thus introducing $\lm \hat \pi_e$ to the Hamiltonian simply adds $\lm$ to the 
energy of the excited state $E_\d \to  \textcolor{black}{E_\d} +\lm$. We may evaluate the amplitudes for the modified Hamiltonian, 
$\h+\lm \hat \pi_e$ and then perform the Fourier transform. We have 
\begin{eqnarray}\label{D1}
A^{\textrm{a+ph}}(e\gets e,t\textcolor{black}{}|\tau) =(2\pi)^{-1}\int_{-\infty}^\infty d\lm \exp[i\lm \t-i(\textcolor{black}{E_\d}+\lm) t -\G t/2]=\n
\exp(-iE_\d t -\G t/2)\delta(\tau -t).\q\q\q\q\q\q\q\q\q\q\q\q 
\end{eqnarray}
Similarly, we find
\begin{eqnarray}\label{D2}
A^{\textrm{a+ph}}(\E\gets e,t\textcolor{black}{}|\tau) 
=
-i\Om\exp(-i\E t) \int_0^t dt'\exp[-i(E_\d-\E) t' -\G t'/2]\delta(\tau-t')\n
=\begin{cases}
-i\Om\exp[-i\E (t-\tau)] \exp[\textcolor{black}{-iE_\d \t-\G \t/2}]\q \text {for}\q 0\le \t \le t\\
0\q\text{otherwise}
\end{cases},
\end{eqnarray}
which is the second of Eqs.(\ref{13}). The remaining  amplitudes are
\begin{eqnarray}\label{D3}
A^{\textrm{a+ph}}(e\gets \E,t\textcolor{black}{}|\tau) =0
\end{eqnarray}
and 
\begin{eqnarray}\label{D4}
A^{\textrm{a+ph}}(\E'\gets \E,t\textcolor{black}{}|\tau) =\exp(-i\E t)\delta(\E-\E')\delta(\t).
\end{eqnarray}
\subsection*{
	Timing the first-order transition in a non-Markovian case}
In the general non-Markovian case, to calculate the required amplitude we
expand, to the first order in $\hat V$, a transition amplitude
\begin{eqnarray}\label{E1}
\la \E|\exp[-i(\h +\lambda \hat \pi_e)t]|e\ra\approx -i\sum_{r'} \Om(E_{r'}) \int_0^t dt'\times \q\q\n
\la \E|\exp[-i(\h_0+\lm  \hat \pi_e)(t-t')]E_{r'}\ra
\la e|\exp[-i(\h_0+\lm  \hat \pi_e)t']|e\ra.
\end{eqnarray}
The integrand reduces to [recall that adding $\lm  \hat \pi_e$ changes  $E_\d$ into $E_e+\lm$ in Eq.(\ref{1b})]
\begin{eqnarray}\label{E2}
\exp[-i\E(t-t')]\delta_{r'r}
\exp[-i(E_e+\lm)t'],
\end{eqnarray}
and performing the Fourier transform with respect to $\lm$ yields
\begin{eqnarray}\label{E3}
A^{\textrm{a+ph}}(\E\gets e,t|\tau) 
=\begin{cases}
-i\Om(\E)\exp[-i\E (t-\tau)] \exp[-iE_\d \t]\q \text {for}\q 0\le \t \le t\\
0\q\text{otherwise}
\end{cases}.\q
\end{eqnarray} 
\textcolor{black}{Using Eqs.(\ref{9}), (\ref{11}) and (\ref{E3}) we find} 
\begin{eqnarray}\label{E4}
A^{\textrm{a+ph+clock}}_{\textrm{Bose}}(\E,n\gets e,0)=const\times \int_0^t \exp[-(\t-\tau_n)^2/\Delta t^2+i(\E-E_e)\tau]d\tau.
\end{eqnarray}
For $\Delta t \to 0$ the limits of integration can be extended to $\pm \infty$. Evaluating the Gaussian integral, 
and taking the absolute square then  yields
\begin{eqnarray}\label{E5}
P(\E,n\gets e,0,t)\approx \frac{\pi \Om^2(\E) \Delta t^2}{[2\pi n]^{1/2}} \exp[-(\E-E_e)^2\Delta t^2/2].
\end{eqnarray}
Replacing ($n_{\textrm{max}}=\om^2t^2N$) the sum $\sum_{n=0}^{n_{\textrm{max}}}n^{-1/2}$ by an integral $\int_0^{n_{\textrm{max}}}n^{-1/2}dn=
2\sqrt{n_{\textrm{max}}}=2 t/\Delta t  $
we obtain the energy distribution of the photons in the presence of a clock
\begin{eqnarray}\label{E6}
P(\E\gets 0,t)= \sum_{n=0}^{n_{\textrm{max}}}P(\E,n\gets e,0,t)\approx {\sqrt {2\pi} \Om^2(\E)\rho(\E) \Delta t} \exp[-(\E-E_e)^2\Delta t^2/2]\times t.
\end{eqnarray}
\subsection*{
	The \e{jump time} in Eq.(\ref{6e})}
Let the decay occur into a finite energy range 
$\Delta E_r =E_{\textrm{max}}-E_{\textrm{min}}$ around $E_\d$,  and assume that $\rh(E_r)\Om^2(E_r)= const$ inside the range, and vanishes 
outside it.
\textcolor{black}{Using Hamiltonian (\ref{1b})}, for the Zeno time we have
\begin{eqnarray}\label{F1}
\t_{\textrm{z}}^2\equiv \left [\la e|\h^2|e\ra -\la e|\h|e\ra^2\right ]^{-1} = [\rh \Om^2 \Delta E_r]^{-1}.
\end{eqnarray}
Recalling that $\G_{\textrm{Fermi}}=2\pi \rh(E_\d)|\la \E=E_e |\h|E_e\ra|^2=2\pi \rh \Om^2$
shows that Eq.(\ref{6e}) reduces to
\begin{eqnarray}\label{F2}
\t_{\textrm{J}} \approx 2\pi/\Delta E_r  \sim 1/\Delta E_r. 
\end{eqnarray}


\section*{Acknowledgements}
DS acknowledges financial support by the Grant PID2021-126273NB-I00 funded by MICINN/AEI/10.13039/501100011033 and by \e{ERDF A way of making Europe}, as well as by the Basque Government Grant No. IT1470-22.

AU and EA acknowledge the financial support by the Ministerio de Ciencia y Innovaci\'on (MICINN, AEI) of the Spanish Government through BCAM Severo Ochoa accreditation CEX2021-001142-S and PID2019-104927GB-C22 grant, as well as by the Basque Government through 
the BERC 2022-2025 Program, IKUR Program, ELKARTEK Programme (grants KK-2022/00006, KK-2021/00022 and KK-2021/00064).

%



\begin{thebibliography}{10}
	\bibitem{TQ1}   Egusquiza \'I.L., Muga J.G. and  Baute A. D., ``Standard'' Quantum Mechanical Approach
	to Times of Arrival in {\it Time in Quantum Mechanics}, vol.1, ed. by J.G. Muga, R. Sala-Mayato, and \'I.L. Egusquiza, Second edition, (Springer verlag, Berlin, Heidelberg, 2008).
	\bibitem{TQ2} Galapon E. A., Post-Pauli's Theorem Emerging Perspective on Time in Quantum Mechanics in {\it Time in Quantum Mechanics}, vol.2, ed. by J.G. Muga, A.Ruschhaupt, and A.Campo,  (Springer verlag, Berlin, Heidelberg, 2008).
	\bibitem{Swiss} Landsman, A.S. \& Keller, U. Attosecond science and the tunnelling time problem.
	{\it Phys. Rep.} \textbf{547}, 1 (2015).
	\bibitem{DSnat}  Sokolovski, D. \& Akhmatskaya, E. 
	No time at the end of the tunnel. {\it Comm. Phys}. {\bf1}, 1-9 (2018). 
	\bibitem{Schul} Schulman, L.S. Jump time and the passage time: The duration of a quantum transition in {\it Time in Quantum Mechanics}, 
	vol.1, ed. by J.G. Muga, R. Sala-Mayato, and \'I.L. Egusquiza, (Springer verlag, Berlin, Heidelberg, 2008).
	\bibitem{Cat}  Schr\"odinger, E. The present situation in quantum mechanics: a translation of Schr\"odinger's  \e {cat paradox} paper 
	(trans. J.D. Trimmer). {\it Am. J. Phys.} \textbf{124},  323 (1980). 
	\bibitem{Gurv} Gurvitz, S. Does the measurement take place when nobody observes it? {\it Fortschr. Phys.} \textbf{65}, 1600065 (2017).
	\bibitem{DSpit} Sokolovski, D. Path integral approach to space-time probabilities:
	A theory without pitfalls but with strict rules. {\it Phys. Rev. D.} \textbf{87}, 076001 (2013).
	\bibitem{FeynL} Feynman,  R.P., Leighton, R.,  Sands, M.  {\it The Feynman Lectures on Physics III}. 
	(Dover Publications, Inc., New York, 1989), Ch.1: Quantum Behavior.  
	\bibitem{Fermi} Merzbacher, E. {\it Quantum Mechanics (3rd ed.).} (Wiley, John \& Sons, Inc., 1998) 
	\bibitem{binom}Feller, W. {\it An Introduction to Probability Theory and Its Applications (3rd ed).} (New York, Wiley, 1968).
	\bibitem{Bruk}Kofler, J. \& Brukner, \v{C}. Classical World Arising out of Quantum Physics under the Restriction of
	Coarse-Grained Measurements. {\it Phys. Rev. Lett.} \textbf{99}, 180403 (2007).
	\bibitem{DSclass} Sokolovski, D., Brourad, S., Alonso, D. From quantum to classical by numbers. 
	{\it New J. Phys.} \textbf{21}, 123031 (2019).
	\bibitem{FeynC} Feynman, R. P. \textit{The
		Character of Physical Law.} (M.I.T. press, Cambridge, Mass, London, 1985).
	\bibitem{Peres} Peres, A. Relativistic quantum measurements, in {\it Fundamental Problems of Quantum Theory.} Ann. N.Y. Acad. Sci. \textbf{755}, (1995).
	\bibitem{Anti} Kaulakys, B. \& Gontis, V. Quantum anti-Zeno effect. {\it Phys. Rev. A.} \textbf{56}, 1131 (1997).
	\bibitem{Copen} Faye, Jan, "Copenhagen Interpretation of Quantum Mechanics", The Stanford Encyclopedia of Philosophy (Winter 2019 Edition), Edward N. Zalta (ed.), URL = https://plato.stanford.edu/archives/win2019/entries/qm-copenhagen/
	
\end{thebibliography}
\end{document}